\documentclass[12pt, fullpage, a4paper]{article}
\usepackage{graphicx, latexsym}
\usepackage{setspace}
\usepackage{subfigure}
\usepackage{apacite}
\usepackage{amssymb, amsmath, amsthm}
\usepackage{bm}
\usepackage{epstopdf}
\usepackage{rotating} 
\usepackage{apacite}
\usepackage{booktabs} 
\usepackage{multirow} 
\usepackage{pifont}
\title{How to relate potential outcomes: Estimating individual treatment effects under a given specified partial correlation} 
\date{}
\author{Mingyang Cai, Stef van Buuren and Gerko Vink}
\date{}

\begin{document}
\maketitle	
	\begin{abstract}
		In most medical research, the average treatment effect is used to evaluate a treatment's performance. However, precision medicine requires knowledge of individual treatment effects: What is the difference between a unit's measurement under treatment and control conditions? In most treatment effect studies, such answers are not possible, because the outcomes under both experimental conditions are not jointly observed. This makes the problem of causal inference a missing data problem. We propose to solve this problem by imputing the individual potential outcomes under a specified partial correlation (SPC), thereby allowing for heterogeneous treatment effects. We demonstrate in simulation that our proposed methodology yields valid inference for the marginal distribution of potential outcomes. We highlight that the posterior distribution of individual treatment effects varies with different specified partial correlations. This property can be used to study the sensitivity of optimal treatment outcomes under different correlation specifications. In a practical example on HIV-1 treatment data, we demonstrate that the proposed methodology generalises to real-world data. Imputing under the SPC therefore opens up a wealth of possibilities for studying heterogeneous treatment effects on incomplete data and the further adaptation of individual treatment effects. 
	\end{abstract}
	

	\section{Introduction}
	\label{sec:1}
	Heterogeneity of treatment effects across individuals is a significant complication in precision treatment assignment to different persons. The difficulty of evaluating individual treatment effects (ITE) from the observed data is that only one of the potential outcomes is observed for each individual (Rubin, 1974\nocite{rubin1974estimating}; Hernan \& Robins, 2010\nocite{hernan2010causal}). This fundamental problem of causal inference implies that causal inference is essentially a missing data problem (Rubin, 2005\nocite{rubin2005causal}; Peng Ding, 2018\nocite{ding2018causal}). Simply ignoring the missingness in the potential outcomes would only allow for average or homogeneous treatment effects. To allow for the estimation of unobserved heterogeneous treatment effects, we need to solve for the individual missing potential outcomes through multiple imputation. 
	
	Multiple imputation is a popular approach for analysing incomplete datasets but is not yet widely used in causal inference. In multiple imputation the missingness is solved before the data is analysed as if it were completely observed. Imputations for missing values are therein drawn from the corresponding posterior predictive distributions in parallel, resulting in multiple imputed datasets. Then, the statistical inference is obtained for each imputed dataset separately by complete-data analyses. Finally, the multiple analyses are aggregated into a single inference using Rubin's rules \cite[pp. 76]{RubinD1987}, which account for within, and across imputation uncertainty. Because multiple imputation imputes the individual potential outcomes, we can evaluate both the difference between outcomes and the individual treatment effects. 
	
	Studying individual treatment effects receives increasing attention. For example, Lamont et al. (2018)\nocite{lamont2018identification}, and Westreich et al. (2015)\nocite{westreich2015imputation} discussed the performance of multiple imputation for potential outcomes. Lamont et al. (2018) applied multiple imputation to evaluate the effectiveness of various programs designed to prevent depression among sampled women and provide program recommendations for women out of the sample. However, Lamont et al. (2018) and Westreich et al. (2015) fitted separate imputation models for potential outcomes based on observed covariates, thereby implicitly assuming conditional independence between potential outcomes. This conditional independence assumption is not always valid and cannot be verified from the observed data \cite{rassler2012statistical}. Imbens and Rubin (2015)\nocite{Imbens2015} and Gadbury et al. (2001)\nocite{gadbury2001evaluating} studied the sensitivity of the average treatment effect estimates under violations of the assumption of conditional independence between potential outcomes. They found that distributions of average causal effect under various partial correlations are different. An alternative imputation strategy fits a fully conditional model for the incomplete outcome. However, Van Buuren (2018)\nocite{Buuren2018} demonstrated that without the specification of the partial correlation, the derived imputations for such models are unstable and can be implausible. 
	
	Specification of the partial correlation in applications of multiple imputation for potential outcomes has received little attention to date. We know that the partial correlation could be an arbitrary value between -1 and 1 in each imputed dataset. However, the imputations become poor when the partial correlation in the imputed dataset is negative (Van Buuren, 2018, section 8.4.1). We therefore assume that this correlation is non-negative. Smink (2016) \nocite{smink2016towards} proposed a data augmentation approach, where rows are added to the data that hold prior information for the partial correlation. This procedure is also outlined in Van Buuren (2018, section 8.4.2)\nocite{Buuren2018}. In Smink's scenario, the imputations are guided by the specified correlation in the augmented cases, but the data does not hold any covariates. One could imagine that augmenting the data with a joint prior set becomes increasingly challenging when the number of covariates increases. 
	
	We propose a new hybrid imputation approach for imputing potential outcomes under a given partial correlation that allows for the collection of incomplete covariates. The procedure is hybrid in the sense that it combines properties from joint modeling imputation (the potential outcomes form a joint and are imputed as such) and fully conditional specification, wherein the covariates are imputed on a fully conditional variable-by-variable basis. In this manuscript we first outline the role of the partial correlation in causal inference, then give a brief overview of multiple imputation and introduce our new hybrid imputation approach. We evaluate the validity of the methodology in simulation and demonstrate the real-world applicability on a clinical trial aimed to evaluate the individual treatment effects of two different therapies on slowing the progression of HIV disease.	
	
	\section{The role of partial correlation}
	\label{sec:2}
	\subsection{Notation}
	\label{sec:2.1}
	Let ${Y(j), j = 1, \dots, p}$ denote one of $p$ incomplete variables and $Y(-j) = (Y(1), \dots, Y(j-1), Y(j+1), \dots, Y(p))$ denote the collection of the $p-1$ variables in $Y$ except $Y(j)$. In this paper, $Y(j)$ ususally represents the potential outcomes. Let $X = (X(1), \dots, X(k))$ be a set of $k$ completely observed variables. Let $\rho(Y(0), Y(1))$ be the correlation between two potential outcomes and $\rho_{Y(0), Y(1)\,|\ X}$ be the partial correlation between two potential outcomes. 
	\subsection{Setup} 
	\label{sec:2.2} 
	We focus on the case of a binary treatment $W_{i}$ and a continuous outcome $Y_{i}$ and assume the data come from a random sample of individuals, indexed by $i \in {1,\dots,N}$. Each individual $i$ has a nonzero probability to be assigned to both treatments, with $W_{i} = 1$ for the active treatment and $W_{i} = 0$ for the control treatment. The number of units under treatment and control are $N_{1}$ and $N_{0}$ respectively. We assume that the treatment assignment mechanism is unconfounded by the unobserved outcomes $Y_{mis}$, i.e., an ignorable assignment mechanism. We also assume that the potential outcomes for any individual are independent of the treatments assigned to others, which is known as the \emph{stable unit treatment value assumption} \cite{Imbens2015}. Here the ignorable or unconfounded assignment mechanism implies that $P(W|Y(0), Y(1), X) = P(W|Y_{obs}, X)$, where X are observed covariates not influenced by treatment assignment. The individual treatment effect is defined as $\tau_{i} = Y_{i}(1) - Y_{i}(0)$. We assume a joint distribution for the potential outcomes $Y_{1}$ and $Y_{0}$ and that the correlation between the potential outcomes $\rho(Y_{1}, Y_{0})$ can be quantified by one or more parameters. The imputation models for missing outcomes $Y(1)$ and $Y(0)$ are: 
	\begin{align}
		\dot{Y}(1) &\sim P(Y^{mis}(1)|Y^{obs}(1), Y(0), X, \dot{\phi}_{1})\\
		\dot{Y}(0) &\sim P(Y^{mis}(0)|Y^{obs}(0), Y(1), X, \dot{\phi}_{0}),
	\end{align}
	where parameters of the imputation model $\dot{\phi}_{1}$ and $\dot{\phi}_{0}$ are draws from their respective posterior distribution.
	\subsection{Partial correlation between potential outcomes}
	\label{sec:2.3}
	The necessity of specifying the partial correlation in the process of multiple imputation has been discussed hereinbefore. This section would further illustrate the causality meaning of the partial correlation. We decompose the individual treatment effect $\tau_{i} \in T, i = 1, \dots, N$ via
	\begin{equation}
		\begin{array}{lr}
			\tau_{i} = Y_{i}(1) - Y_{i}(0) = X_{i}^{\mathsf{T}}\beta + \varepsilon_{i},
		\end{array}
	\end{equation}
	where $X_{i}$ are observed pre-treatment covariates for individual $i$. Under random treatment assignment, the regression weight $\beta$ could be the ordinary least-square (OLS) estimation of $\tau_{i}$ on $X_{i}$. The quantity $X_{i}^{\mathsf{T}}\beta$ is known as systematic treatment effect variation and the residual $\varepsilon_{i}$ is the idiosyncratic treatment effect variation not explained by $X_{i}$ (Ding et al., 2019\nocite{ding2019decomposing}; Heckman et al., 1997; Djebbari \& Smith, 2008\nocite{djebbari2008heterogeneous}). The idiosyncratic variation accounts for treatment effect variation not attributable to differences in observed covariates. Based on the formula of OLS estimation, the coefficient 
	\begin{equation}
		\begin{array}{cl}
			\beta &= (X^{\mathsf{T}}X)^{-1}X^{\mathsf{T}}T \\
			&= (X^{\mathsf{T}}X)^{-1}X^{\mathsf{T}}Y(1) - (X^{\mathsf{T}}X)^{-1}X^{\mathsf{T}}Y(0) \\
			&= \beta_{1} -\beta_{0},
		\end{array}
	\end{equation}
	where $\beta_{1}$ and $\beta_{0}$ are the corresponding regression weights of the potential outcomes $Y_{1}$ and $Y_{0}$ on the observed covariates $X$. Similarly, the idiosyncratic treatment effect variation $\varepsilon_{i}$
	\begin{equation}
		\begin{array}{cl}
			\varepsilon_{i} &= \tau_{i} - X_{i}^{\mathsf{T}}\beta\\
			&= (Y_{i}(1)  - X_{i}^{\mathsf{T}}\beta_{1}) - (Y_{i}(0) - X_{i}^{\mathsf{T}}\beta_{0})\\
			&= \varepsilon_{i}(1) -\varepsilon_{i}(0),
		\end{array}
	\end{equation} 
	where $\varepsilon(1)$ and $\varepsilon(0)$ are the residuals from the regression of the potential outcomes $Y_{1}$ and $Y_{0}$ on the observed covariates $X$. Applying the theory of variance decomposition for linear regression, we could decompose the variance of individual treatment effect into two components:
	\begin{equation}
		\text{Var}(\tau_{i}) = \text{Var}(X_{i}^{\mathsf{T}}\beta) + \text{Var}(\varepsilon_{i}). 
	\end{equation}
	Based on the idiosyncratic treatment effect variation formula (5), the idiosyncratic components of individual treatment variance becomes
	\begin{equation}
		\begin{array}{cl}
			\text{Var}(\varepsilon_{i}) &= \text{Var}[\varepsilon_{i}(1) -\varepsilon_{i}(0)]\\ 
			&= \text{Var}[\varepsilon_{i}(1)] + \text{Var}[\varepsilon_{i}(0)] - 2\text{Cov}[\varepsilon_{i}(1), \varepsilon_{i}(0)], 
		\end{array}
	\end{equation} 
	which demonstrate that partial correlation $\rho_{Y(0)Y(1)\,|\ X}$ between the potential outcomes does impact the idiosyncratic variation, which is unidentifiable from the observed data. 
	
	Although the relation between potential outcomes cannot be determined solely from the observed data, there are still some approaches to identifying the partial correlation, such as the model-based approach, the experiment-based approach, and sensitivity analysis. The model-based approach explicitly models the relationship between the idiosyncratic variation and the assignment mechanisms such that the partial correlation would be close or equal to zero \cite{heckman2005scientific}. Economists usually deal with ex-post causal inference. The agents select the treatment according to their ex-ante evaluation. In this case, economists could infer the information only available to agents from the assignment mechanism.  For instance, Heckman (2010)\nocite{heckman2010effects} investigated the causal effect of educational decisions on the labour market and health outcomes. He modelled latent cognitive and social-emotional endowments and included these latent variables into the outcome equations. As indicated earlier, the partial correlation between potential outcomes is the only unknown parameter in the formula of idiosyncratic variation. Therefore, the model for idiosyncratic variation could be tailored to the model for the partial correlation between potential outcomes. Generally, the model-based approach attempts to figure out latent variables affecting the partial correlation between potential outcomes. 
	
	The experiment-based approach intends to design sophisticated experiments to collect additional data where analysts could evaluate the partial correlation between potential outcomes. For instance, the experiment-based approach collects repeated measurements under more than one treatment level for the same individual from which some relevant information about the partial correlation between potential outcomes is available. The experiment designed for repeat measurements is known as N-1 trails(Shamseer et al., 2015\nocite{shamseer2015consort}; Araujo et al., 2016\nocite{araujo2016understanding}). Researchers could also design an auxiliary treatment ($W_{i} = 2$) and assign the extensive treatment to all individuals in the sample. Then the individual treatment effect $Y_{i}(1) - Y_{i}(0)$ can be evaluated by $[Y_{i}(2) - Y_{i}(0) - (Y_{i}(2) - Y_{i}(1))]$. 
	
	Both model-based and experiment-based approaches search for extra information to determine the partial correlation between potential outcomes. If such additional information is not available, the alternative is sensitivity analysis. After imputing the missing potential outcomes, one could evaluate individual treatment effects with various valid partial correlations and study the effect of partial correlations on the conclusions (Gadbury, Iyer \& Allison, 2001)\nocite{gadbury2001evaluating}.  
	
	\section{Multiple imputation of multivariate incomplete variables}
	\label{sec:3}
	Datasets used for evaluating individual treatment effects by multiple imputation often have incomplete covariates and potential outcomes. We impute potential outcomes with joint modeling (JM) and covariates with fully conditional specification (FCS). Usually, one framework is used to generate all imputations, but Van Buuren (2018) highlighted that a blocked approach could be adopted to accommodate for hybrid versions of JM within FCS. We will now briefly introduce joint modelling, fully conditional specification and so-called hybrid imputation. 
	
	\subsection{Joint modeling imputation (JM)}
	\label{sec:3.1}
	Joint modeling imputation assumes a model $p(Y^{mis}, Y^{obs}\,|\,\theta)$ for the complete data and a prior distribution $p(\theta)$ for the parameter $\theta$. Joint modelling partitions the observed data into groups based on the missing pattern and imputes the missing data within each missing pattern according to corresponding predictive distribution. Under the assumption of ignorability, the parameters of the predictive distribution for different missing patterns are generated from the posterior joint distribution. Schafer (1997) proposed joint modelling methods for multivariate normal data, categorical data and mixed normal-categorical data. The joint modelling approach has solid theoretical properties (i.e., compatibility between the imputation and substantive models) while it lacks the flexibility of model specification.
	
	\subsection{Fully Conditional Specification (FCS)}
	\label{sec:3.2}
	In Fully Conditional Specification, we specify the distribution for each partially observed variable conditional on all other variables $P(Y(j)|Y(-j), X, \theta_{j})$ and impute each missing variable iteratively. The FCS starts with naive imputations such as a random draw from the observed values. The \emph{t}th iteration for the incomplete variable \emph{$Y(j)$} consists of the following draws:
	\begin{align*}
		&\theta_{j}^{t} \sim f(\theta_{j})f(Y^{obs}(j)|Y^{t-1}(-j), X, \theta_{j})\\
		&Y^{mis(t)}(j) \sim f(Y^{mis}(j)|Y^{t}(-j), X, \theta_{j}^{t}),
	\end{align*}
	where $f(\theta_{j})$ is generally specified as a noninformative prior. After a sufficient number of iteration, typically with 5 to 10 iterations \cite{Buuren2018}, the stationary distribution is achieved. The final iteration generates a single imputed dataset and the multiple imputations are created by applying FCS in parallel \emph{m} times. Since FCS provides tremendous flexibility in specifying imputation models for multivariate partially observed data, FCS is now a widely accepted and popular MI approach \cite{van2007multiple}. Even while, FCS lacks a satisfactory theory and has a potential risk in incompatibility. 
	
	\subsection{Block imputation}
	\label{sec:3.3}
	Block imputation combines the flexibility of FCS with the attractive theoretical properties of JM. A block consists of one or more variables. If the block has multiple variables, then we use multivariate imputation methods to impute those variables jointly. A simple example would be multiple imputation of missing variables with quadratic effects $Y = \alpha + \beta_{1}X + \beta_{2}X^2 + \varepsilon$. In such a case, grouping the missing variable $X$ and its corresponding square term $X^2$ within one block is of benefit to preserve the quadratic relationship (Vink, 2019)\nocite{vink2019}. The joint modelling approach is the special case where all variables form one block, while the FCS approach treats each variable as a separate block. 
	
	When the imputation model of one variable is potentially incompatible, or its theoretical properties are not fully studied (i.e., whether the imputations based on the FCS correspond to drawing from a joint distribution), block imputation would merge that variable with other variables and apply the joint modeling imputation approach to that block. On the other hand, when the joint distribution of several missing variables is ambiguous, block imputation could use the FCS approach to impute each variable. In general, the apparent advantage of block imputation is the flexibility of model specification. However, block methods are hardly known or studied. While available in the \texttt{mice} software, the properties of block imputation have yet to be studied. 
	
	\section{Specified partial correlation imputation}
	\label{sec:4}
	\begin{figure}[t!]
		\centering
		\includegraphics[width=1.0\linewidth,height=0.5\textheight]{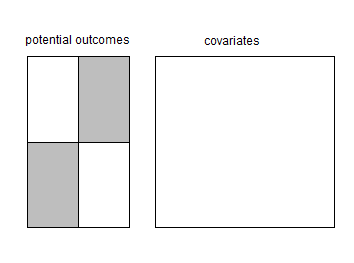} 
		\caption{Missingness mechanism of potential outcomes. The write represents observed value and the grey represents missing value. Without loss of generality, we assume covariates are completely observed.}
		\label{fig1}
	\end{figure}
	
	In this section we detail how blocked imputation can be used to impute missing outcomes with a given partial correlation between potential outcomes. We term the algorithm the imputation algorithm with specified partial correlation (SPC). Since the missing pattern of potential outcomes is somehow restrictive (see Figure \ref{fig1}) that is, no cases with completely observed potential outcomes, the imputation procedure follows three steps: 
	
	\begin{enumerate}
		\item Estimating the marginal distribution of potential outcomes conditioning on pre-treatment variables. 
		\item Derive the multivariate density of potential outcomes by combining the marginal distribution of potential outcomes and the specified correlation between potential outcomes. 
		\item Impute the missing outcomes with the corresponding submodel obtained from the multivariate distribution. 
	\end{enumerate}	
	
	Based on Rubin causal model (Imbens \& Rubin 2015, Ch 8), it is plausible to assume a multivariate normal distribution for continuous potential outcomes. However, it is usually not valid to assume a joint distribution for the incomplete dataset. Applying fully conditional specification to impute covariates allows flexibility of imputation model specification. It is noticeable that SPC could also predict the posterior distribution of the individual treatment effect for units not in the experiment.  
	
	Our approach shares some similarities with statistical matching discussed by Moriarity \& Scheuren (2003)\nocite{moriarity2003note}. For example, suppose there are two sample files, A and B. File A collects variables $X$ and $Y$ and file B collects variables $X$ and $Z$. The purpose of statistical matching is to combine two files, A and B, into one file containing variables $X$, $Y$ and $Z$. Rubin (1986)\nocite{rubin1986statistical} proposed a procedure of statistical matching with three steps: regression step, matching step and concatenation step. In the regression step, Rubin specified the correlation between variable $X$ and $Y$ to derive the joint distribution of ($X$, $Y$, $Z$) in two sample files. We develop this idea to evaluate the individual treatment effects and extend to multiple treatments condition.  
	
	Without loss of generality, let us assume that potential outcomes follow a multivariate normal distribution. We specify Bayesian linear models for two potential outcomes based on observed covariates. 
	\begin{align}
		Y(0) = \beta_{0}X + \varepsilon_{0}, \varepsilon_{0} \sim \mathcal{N}(0, \sigma_{0}^2)\\
		Y(1) = \beta_{1}X + \varepsilon_{1}, \varepsilon_{1} \sim \mathcal{N}(0, \sigma_{1}^2).
	\end{align}
	Bayesian sampling draws $\beta_{0}^{*}$, $\beta_{1}^{*}$, $\sigma_{0}^{*2}$, $\sigma_{1}^{*2}$ from their respective posterior distribution. The Jeffrey's prior used and hence, the posterior distributions of $\sigma_{0}^{*2}$ and $\sigma_{1}^{*2}$ would be inverse $\chi^2$ distribution:
	\begin{align}
		\sigma_{0}^{*2} \sim \sum_{i = 1}^{N_{0}}(Y_i(0) - \hat{\beta}_{0}X_{i})^2\chi_{N_{0} - k}^{-2}\\
		\sigma_{1}^{*2} \sim \sum_{i = 1}^{N_{1}}(Y_i(1) - \hat{\beta}_{1}X_{i})^2\chi_{N_{1} - k}^{-2},
	\end{align}
	where $\hat{\beta}_{0} = (X_{0}'X_{0})^{-1}X_{0}'Y(0)$, $\hat{\beta}_{1} = (X_{1}'X_{1})^{-1}X_{1}'Y(1)$ and \emph{k} is the number of covariates. The conditional distributions of  $\beta_{0}^{*}$ and $\beta_{1}^{*}$ are multivariate normal:
	\begin{align}
		\beta_{0}^{*} | \sigma_{0}^{*2} \sim \mathcal{N}(\hat{\beta}_{0}, \sigma_{0}^{*2}((X_{0}'X_{0})^{-1}))\\
		\beta_{1}^{*} | \sigma_{1}^{*2} \sim \mathcal{N}(\hat{\beta}_{1}, \sigma_{1}^{*2}((X_{1}'X_{1})^{-1})).
	\end{align}
	Since there is no information relevant to the partial correlation between potential outcomes in the observed data, the posterior distribution of the partial correlation $\rho_{Y(0)Y(1)|X}$ equals the prior distribution specified by the user, who can select several numbers in the interval [$-1$, $1$] to investigate the sensitivity to $\rho_{Y(0)Y(1)|X}$. 
	Finally, combined the marginal distribution of $Y(0)$ and $Y(1)$ with the specification of $\rho_{Y(0)Y(1)|X}$, the joint distribution of $Y^{mis}(0)$ and $Y^{mis}(1)$ is:
	\begin{eqnarray}
		\left[\begin{array}{c}
			Y^{mis}(0)\\
			Y^{mis}(1)
		\end{array}\right] & \sim & \mathcal{N}\left[\left(\begin{array}{c}
			\beta^{*}_{0}x\\
			\beta^{*}_{1}x
		\end{array}\right),\left(\begin{array}{cc}
			\sigma^{*2}_{0} & \rho_{Y(0)Y(1)|X}\sigma^{*}_{0}\sigma^{*}_{1}\\
			\rho_{Y(0)Y(1)|X}\sigma^{*}_{0}\sigma^{*}_{1} & \sigma^{*2}_{1}
		\end{array}\right)\right],
	\end{eqnarray}
	and the distributions of $Y^{mis}(0)$ and $Y^{mis}(1)$ are:
	\begin{align}
		Y^{mis}(0) \sim \mathcal{N} (\beta^{*}_{0}x + (Y(1) - \beta^{*}_{1}x)\rho_{Y(0)Y(1)|X}\sigma_{0} / \sigma_{1}, (1 - \rho_{Y(0)Y(1)|X}^2)\sigma^{*2}_{0})\\
		Y^{mis}(1) \sim \mathcal{N} (\beta^{*}_{1}x + (Y(0) - \beta^{*}_{0}x)\rho_{Y(0)Y(1)|X}\sigma_{1} / \sigma_{0}, (1 - \rho_{Y(0)Y(1)|X}^2)\sigma^{*2}_{1}).
	\end{align}
	Comparing equations (15) and (16) to equations (8) and (9), it is evident that inclusion of the observed outcome may change the location of missing outcomes shifts slightly and the uncertainty is reduced when imputing missing outcomes under the specified correlation between potential outcomes. For the prediction of units out of trials, the reasonable values for outcomes under two treatments could be drawn from the joint distribution (14). 
	
	When generalizing to the multiple treatments condition $W = 0, 1,\dots,w$, the marginal posterior distribution for potential outcomes would be:
	\begin{align*}
		Y(0)^{*} &= \beta_{0}^{*}X + \varepsilon_{0}^{*}, \varepsilon_{0}^{*} \sim \mathcal{N}(0, \sigma_{0}^{*2})\\
		Y(1)^{*} &= \beta_{1}^{*}X + \varepsilon_{1}^{*}, \varepsilon_{1}^{*} \sim \mathcal{N}(0, \sigma_{1}^{*2})\\
		&\dots\\
		Y(w)^{*} &= \beta_{w}^{*}X + \varepsilon_{w}^{*}, \varepsilon_{w}^{*} \sim \mathcal{N}(0, \sigma_{w}^{*2}),
	\end{align*}
	where the values of $\beta_{0}^{*}$, $\beta_{1}^{*}$, \dots, $\beta_{w}^{*}$, $\sigma_{0}^{*2}$, $\sigma_{1}^{*2}$, \dots, $\sigma_{w}^{*2}$  draw from their respective Bayesian posterior distribution.
	If $\sigma_{0}^{*2}, \dots, \sigma_{w}^{*2}$ are unrestricted, with pairwise specification of partial correlation between potential outcomes, the joint distribution of $Y^{mis}(0)$, $Y^{mis}(1)$, \dots, $Y^{mis}(w)$ is:
	\begin{eqnarray}
		\left[\begin{array}{c}
			Y^{mis}(0)\\
			Y^{mis}(1)\\
			\dots\\
			Y^{mis}(w).
		\end{array}\right] & \sim & \mathcal{N}(\mathcal{M}, \Sigma),
	\end{eqnarray}
	where  $\mathcal{M} = (\beta^{*}_{0}x, \beta^{*}_{1}x, \dots, \beta^{*}_{w}x)^{\mathsf{T}}$. The covariance matrix $\Sigma$ must be positive semi-definite:
	\begin{eqnarray}
		\begin{array}{c}
			\Sigma
		\end{array} = \left(\begin{array}{cccc}
			\sigma^{*2}_{0} & \rho_{Y(0)Y(1)|X}\sigma^{*}_{0}\sigma^{*}_{1} & \dots & \rho_{Y(0)Y(w)|X}\sigma^{*}_{0}\sigma^{*}_{w}\\
			\rho_{Y(1)Y(0)|X}\sigma^{*}_{1}\sigma^{*}_{0} & \sigma^{*2}_{1} & \dots & \rho_{Y(1)Y(w)|X}\sigma^{*}_{1}\sigma^{*}_{w}\\
			\vdots & \vdots & \ddots & \vdots\\
			\rho_{Y(w)Y(0)|X}\sigma^{*}_{w}\sigma^{*}_{0} & \rho_{Y(w)Y(1)|X}\sigma^{*}_{w}\sigma^{*}_{1} & \dots & \sigma^{*2}_{w}
		\end{array}\right).
	\end{eqnarray} 
	Draws of missing outcomes for units under different treatments could be derived from the joint distribution based on the property of conditional distribution for the multivariate normal distribution. For instance, with units under control treatment $W = 0$, the distribution of missing outcomes $Y^{mis}(-0) = (Y^{mis}(1), \dots, Y^{mis}(w))$ would be $Y^{mis}(-0) \sim \mathcal{N} ((\beta^{*}_{1}x, \dots, \beta^{*}_{w}x)^{\mathsf{T}} + \Sigma_{0-0}\Sigma_{-0-0}^{-1}(Y_0 - \beta^{*}_{0}x), \Sigma_{00} - \Sigma_{0-0}\Sigma_{-0-0}^{-1}\Sigma_{-00})$, where
	\begin{eqnarray}
		\left[\begin{array}{cc}
			\Sigma_{00} & \Sigma_{0-0}\\
			\Sigma_{-00} & \Sigma_{-0-0}
		\end{array}\right],
	\end{eqnarray}
	is the partition of $\Sigma$: $\Sigma_{00} = \sigma^{*2}_{0}$, $\Sigma_{0-0} = \Sigma_{-00}^{\mathsf{T}} = (\rho_{Y(0)Y(1)|X}\sigma^{*}_{0}\sigma^{*}_{1}, \dots, \rho_{Y(0)Y(w)|X}\sigma^{*}_{0}\sigma^{*}_{w})$ and 
	\begin{eqnarray}
		\begin{array}{c}
			\Sigma_{-0-0}
		\end{array} = \left(\begin{array}{cccc}
			\sigma^{*2}_{1} & \dots & \rho_{Y(1)Y(w)|X}\sigma^{*}_{1}\sigma^{*}_{w}\\
			\vdots & \vdots & \ddots & \vdots\\
			\rho_{Y(w)Y(1)|X}\sigma^{*}_{w}\sigma^{*}_{1} & \dots & \sigma^{*2}_{w}
		\end{array}\right).
	\end{eqnarray} 
	One could use the sweep operator for rapid calculation of the parameters for imputation models of missing outcomes \cite{goodnight1979tutorial}. 
	
	\section{Simulation study}
	\label{sec:5}
	We evaluate the performance of SPC at both the individual level (i.e. the individual treatment effect) and the aggregate level (i.e. the average treatment effect). For individual causal inference, we study the mean and mean absolute differences between the `true' and imputed individual treatment effects, together with posterior distributions of individual treatment effects. For average causal inference, we analyse biases and confidence interval coverages of the estimated parameters in the distribution of the potential outcomes. We perform a sensitivity analysis to the multiple imputation approach with three different values for the partial correlation between potential outcomes: $\rho = 0, 0.73 $ or $0.99$, which correspond to, respectively, a conditional independent correlation assumption, the correct partial correlation and a constant treatment effect condition. 
	
	We compare the performance of SPC to the targeted learning approach by Van der Laan \& Rose (2011). Targeted learning is an alternative for estimating individual treatment effects. The idea is to estimate the data-generating distribution $P_{0}$ and then update the initial estimation to make an optimal bias-variance tradeoff for the scientific interest $\Psi(P_{0})$. To estimate individual treatment effects, we define the average treatment effect as the scientific interest:
	\begin{equation}
		\Psi(P_{0}) = E[E(Y_{i}(1)\,|\,X) - E(Y_{i}(0)\,|\,X)].
	\end{equation} 
	The targeted learning consists of three steps of analysis: 1) definition of the data-generating model and the scientific interest $\Psi(P_{0})$, 2) super learning for initial prediction of $\Psi(P_{0})$ and 3) targeted maximum likelihood estimation for $\Psi(P_{0})$ \cite{van2011targeted}. Specifically, before estimation, it is necessary to define a set of possible probability distributions of observed data and identify a collection of causal assumptions (i.e., the ignorable assignment mechanism and the stable unit treatment value assumption) to the identification of the correct model. With the definition of the model, one could apply the super learner to derive an initial estimation for the distribution of potential outcomes $\hat{P}_0$. The super learner first selects a library of candidate algorithms and a risk function and then applies the validation set approach to calculate the average risk for each algorithm. The optimal algorithm with the smallest average risk is used to produce the initial predicted distribution of potential outcomes. The candidate algorithms could be parametric(i.e., general linear model), non-parametric (i.e., random forest), or even a weighted combination of statistical algorithms. 
	
	After the initial estimation of predicted potential outcomes, one could define the targeted maximum likelihood estimation (TMLE) for scientific interest $\Psi(P_{0})$. The TMLE step reduces the bias in the estimation of $\Psi(P_{0})$ if the initial estimation $\Psi(\hat{P}_0)$ is inconsistent. This is accomplished by exploiting information in the treatment assignment mechanisms to adjust the initial estimations. Generally, the adjustment is an iterative procedure. However, when the scientific interest is the average treatment effect, convergence is achieved in one step. More details are provided in Gruber \& van der Laan (2011)\nocite{gruber2011tmle}.    
	
	While targeted learning is a machine learning approach aimed at estimating the average treatment effect, it involves calculating the missing potential outcomes and can therefore also be used to identify individual treatment effects. However, unlike SPC, the targeted learning fits the distribution of the missing outcome only based on the covariates, which assumes conditional independence between potential outcomes. In the simulation study, we aim to show the relevance of specifying the correlation between potential outcomes when specifying the analysis model of potential outcomes. 
	
	\subsection{Simulation conditions}
	\label{sec:5.1}
	We design the two potential outcomes $Y(0)$ and $Y(1)$ as well as one baseline covariate $X$. The data is generated with a multivariate normal distribution:
	\begin{eqnarray*}
		\begin{pmatrix}Y_{0}\\
			Y_{1}\\
			X
		\end{pmatrix} & \sim & \mathcal{N}\left[\left(\begin{array}{c}
			0\\
			1\\
			2
		\end{array}\right),\left(\begin{array}{ccc}
			1 & 0.8 & 0.5\\
			0.8 & 1 & 0.5\\
			0.5 & 0.5 & 1
		\end{array}\right)\right]\\
	\end{eqnarray*}
	Because the marginal correlation between potential outcomes is 0.8, the corresponding partial correlation is 0.73. 
	\begin{equation}
		\begin{array}{cl}
			\rho_{y_{0}y_{1}|x} &= \frac{\rho_{y_{0}y_{1}} - \rho_{y_{0}x}\rho_{y_{1}x}}{\sqrt{1 - \rho_{y_{0}x}^2}\sqrt{1 - \rho_{y_{1}x}^2}} \\
			&= \frac{0.8 - 0.25}{0.75}\\
			&\approx 0.73,
		\end{array}
	\end{equation}
	A total of $N = 5000$ independent and identically distributed cases are generated. The first 2500 cases have only information for $Y_0$ and the remaining cases have only information on $Y_1$. One thousand repetitions of the simulation are produced for average causal inference. While, for individual causal inference, we derive the posterior distributions of imputed outcomes from twenty imputed datasets. For reasons of brevity, we only include one pre-treatment covariate. However, it is straightforward to extend the methodology to situations with more covariates and even mixtures of continuous and categorical predictors.
	
	\subsection{Results}
	\label{sec:5.2}
	\subsubsection{Individual causal inference}
\begin{figure}[ht!]
		\begin{center}
			\resizebox{\textwidth}{!}{
				\subfigure[SPC $\rho_{partial} = 0$]{
					\label{boxplot:a}
					\includegraphics[scale=.5]{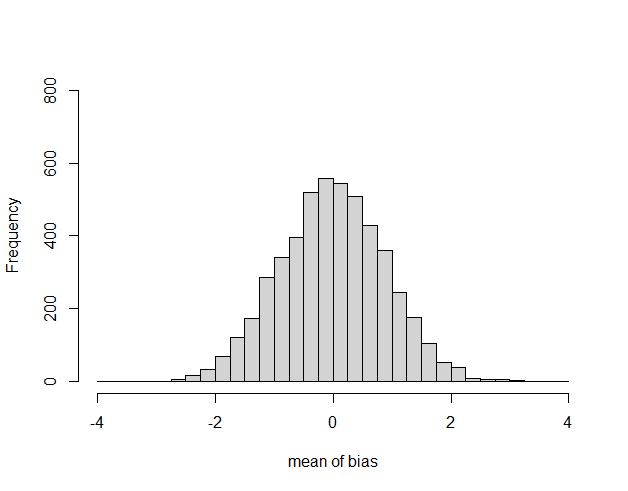}
				}
				\subfigure[SPC $\rho_{partial} = 0.73$]{
					\label{boxplot:b}
					\includegraphics[scale=.5]{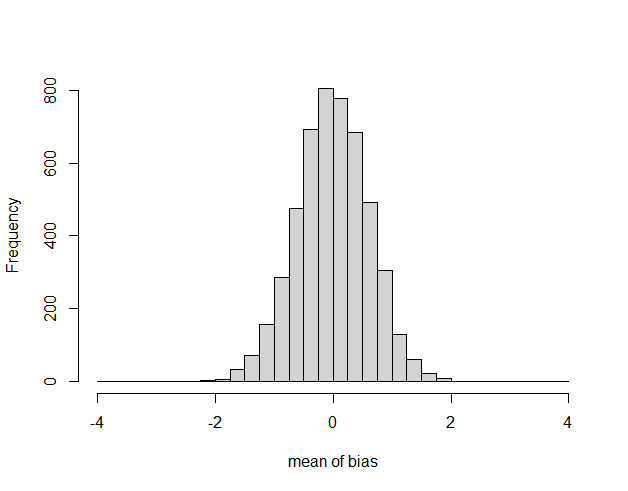}
				}
			}\\ 	
			\resizebox{\textwidth}{!}{
				\subfigure[SPC $\rho_{partial} = 0.99$]{
					\label{boxplot:c}
					\includegraphics[scale=.5]{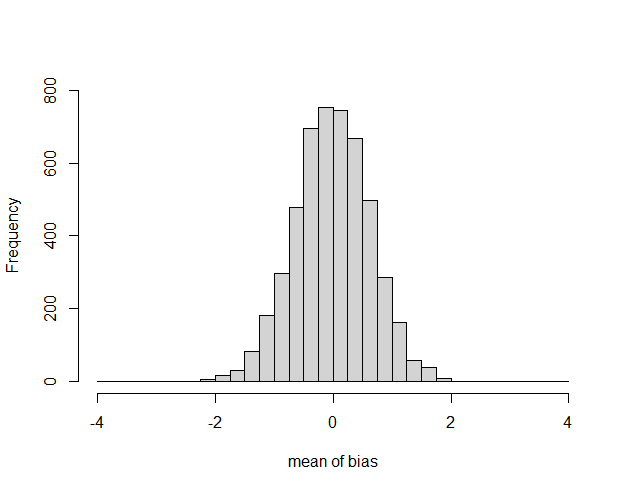}
				}
				\subfigure[The targeted learning]{
					\label{boxplot:d}
					\includegraphics[scale=.5]{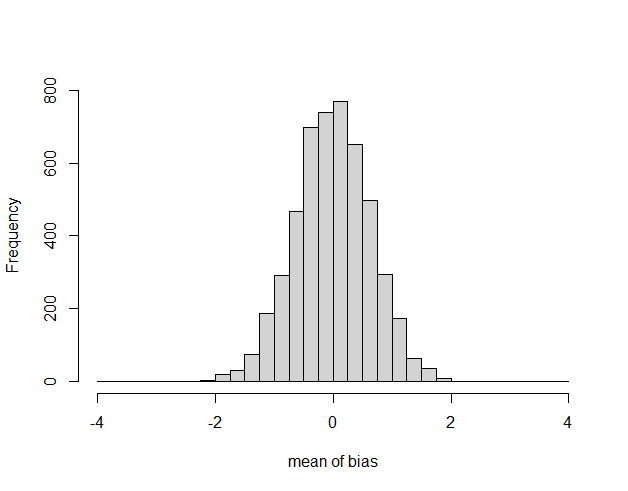}
				}
			}
		\end{center}
		\caption{Histrogram plots of the bias of estimate the individual treatment effects for all four strategies.}
		\label{fig2}
	\end{figure}

	We define bias as the mean difference between \emph{true} and estimated values over twenty imputed datasets. Figure \ref{fig2} shows the distribution of the bias for all four strategies, and the corresponding location and scale are displayed in Table \ref{tab1}.  
	
	\begin{table}[ht!]
		\centering
		\begin{tabular}{c|ccc}
			$\rho_{Y(0)Y(1)\,|\ X}$                    & Mean bias   & Variance of bias \\
			\hline                              
			SPC $\rho_{partial} = 0$       & -0.012     & 0.791 \\
			SPC $\rho_{partial} = 0.73$    & -0.007     & 0.363 \\
			SPC $\rho_{partial} = 0.99$    & -0.013     & 0.398 \\
			The targeted learning         & -0.006 & 0.401 
		\end{tabular}
		\caption{The location and the scale for the bias of estimate the individual treatment effects for all four strategies.}
		\label{tab1}
	\end{table} 
	
	\begin{figure}[ht!]
		\centering
		\includegraphics[width=1.0\linewidth,height=0.5\textheight]{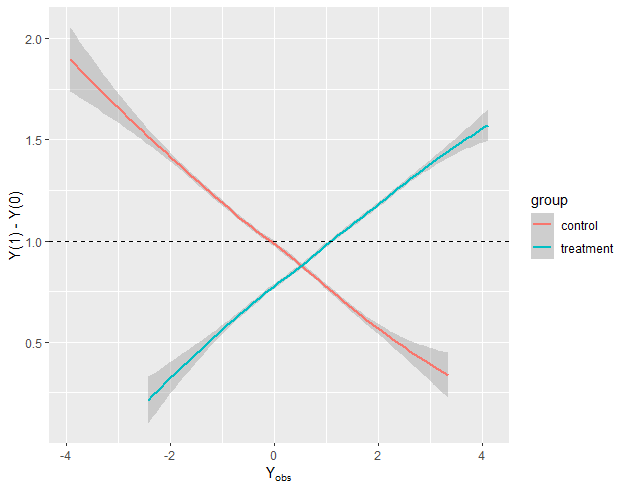} 
		\caption{The plot shows mean individual treatment effects $(Y(1) - Y(0))$ in both treatment and control group when partial correlation is defined as $0.73$. The oucome $Y(0)$ is observed in the control group and the oucome $Y(1)$ is observed in the treatment group. The dashed line represents the average treatment effect.}
		\label{fig3}
	\end{figure}
	
	Overall, SPC with three different partial correlations and the targeted learning all yield unbiased estimates of the average treatment effect. However, in terms of the scale, The SPC with the \emph{correct} partial correlation has a minor variance. The closer the specified partial correlation is to the partial correlation in the \emph{true} data generating model, the smaller bias and variance can be expected. Although it is difficult to produce accurate estimates of the individual treatment effect for units at the tail in Fig \ref{fig3}, we still have a large proportion of the estimated individual treatment effect with negligible biases. Since the variance of the bias equals the partial variance of the potential outcomes, we could include more explanatory variables to increase the accuracy of the imputation of missing outcomes and hence the prediction of individual treatment effect. On the other hand, when the specified partial correlation deviates from the \emph{true} value, more variance would appear because of the differences between the \emph{true} distribution and the estimated distribution for each missing outcome.

	\begin{figure}[ht!]
		\begin{center}
			\resizebox{\textwidth}{!}{
				\subfigure[SPC $\rho_{partial} = 0$]{
					\label{boxplot:e}
					\includegraphics[scale=.5]{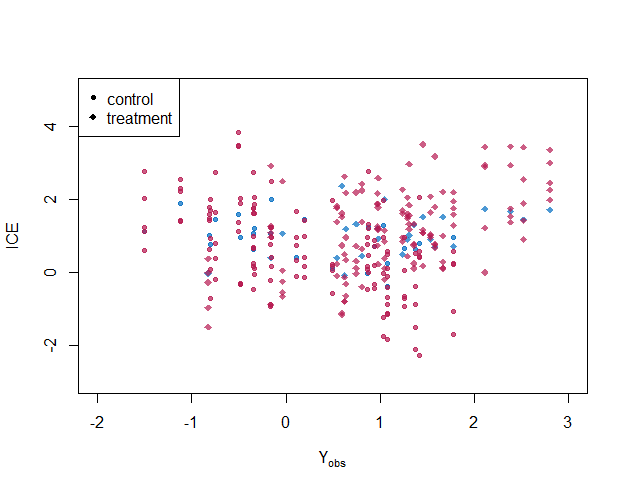}
				}
				\subfigure[SPC $\rho_{partial} = 0.73$]{
					\label{boxplot:f}
					\includegraphics[scale=.5]{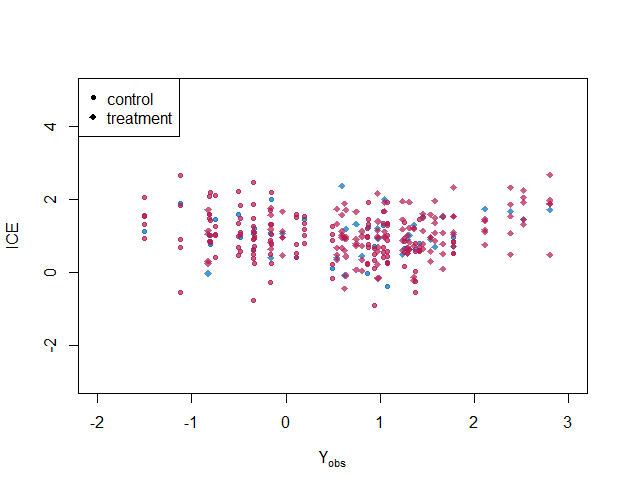}
				}
			}\\ 	
			\resizebox{\textwidth}{!}{
				\subfigure[SPC $\rho_{partial} = 0.99$]{
					\label{boxplot:g}
					\includegraphics[scale=.5]{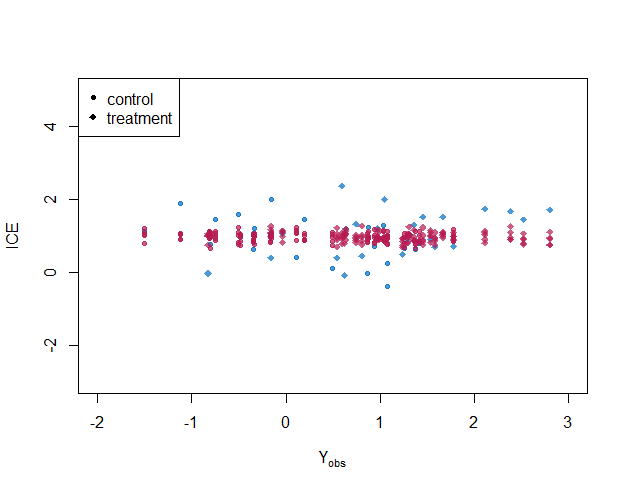}
				}
				\subfigure[The targeted learning]{
					\label{boxplot:h}
					\includegraphics[scale=.5]{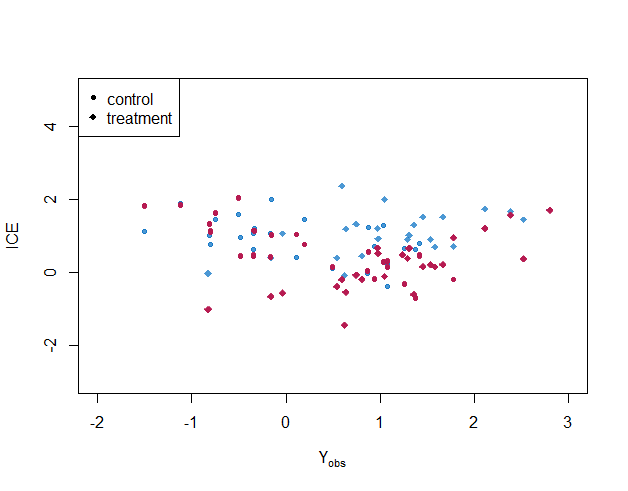}
				}
			}
		\end{center}
		\caption{Stripplot of $\texttt{m} = 5$ of observed (blue) and imputed (red) data for individual treatment effects with selected cases.}
		\label{fig4}
	\end{figure}
	Figure \ref{fig4} shows posterior distributions of individual treatment effects for selected cases($\emph{i} = 100, 200, \dots, 5000$). When the partial correlation is specified correctly, the imputations look plausible: imputed ITE covers the \emph{true} ITE for almost every case, and the variance of ITE for each individual is smaller than the case under the independent conditional correlation assumption. With homogeneous treatment effect assumption, i.e., $\rho_{Y(0)Y(1)\,|\ X} = 0.99$, the imputed individual treatment effects are biased towards the average treatment effect. For targeted learning, uncertainty about the missing outcomes is not estimated. 
	
\begin{figure}[ht!]
		\centering
		\includegraphics{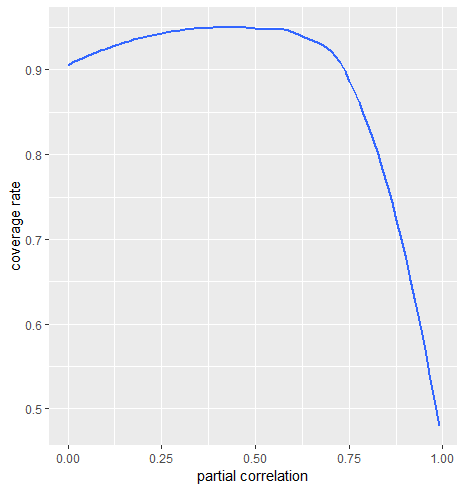} 
		\caption{Coverage rate of the posterior distribution of estimated individual treatment effects}
		\label{fig5}
	\end{figure}

\begin{figure}[ht!]
		\centering
		\includegraphics{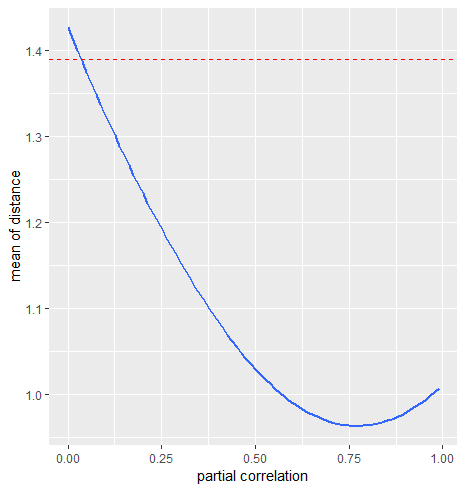} 
		\caption{Mean distance between the \emph{true} and the mean estimated individual treatment effects. The red dashed line represents the mean distance calculated by the targeted learning approach.}
		\label{fig6}
	\end{figure}

	Furthermore, we evaluate the imputations with all possible positive partial correlation (range from 0 to 1) by mean distance between the \emph{true} and the mean estimated individual treatment effects and the rate of the posterior distribution of imputed outcome cover the \emph{true} value. Fig \ref{fig5} shows that the violation of the homogeneous treatment effect assumption leads to extremely poor coverage. If we specified the partial correlation closed to the \emph{true} value, the imputed outcomes would be more accurate (see Fig \ref{fig6}). Fig \ref{fig6} also highlight the mean distance calculated by the targeted learning, which implies the implicit assumption of conditional independence between potential outcomes. 
	
	The SPC approach derives the distribution of individual treatment effects, which provides more information on treatment recommendations. For instance, with a small individual treatment effect, it is possible to estimate the probability of a positive treatment effect from the distribution of individual treatment effects.    
	
	\subsubsection{Average causal inference}
	In this section, we investigate whether we could provide valid inferences for the distribution of the potential outcomes. In addition, we are interested in the biases and the coverage of nominal 95\% confidence intervals of all parameters related to the potential outcomes.  
	\begin{table}[ht!]
		\centering
		\begin{tabular}{lccc}
			Method & Truth & Est   & Cover \\
			\hline
			SPC $\rho_{partial} = 0$       &      &      &  \\\hline
			E($y_0$)       & 0.0     & 0.00     & 0.95 \\
			E($y_1$)       & 1.0     & 1.00     & 0.94 \\
			Var($y_0$)     & 1.0     & 1.00 & 0.94 \\
			Var($y_1$)     & 1.0     & 1.00 & 0.94 \\
			Cov($y_0$, $y_1$) & 0.8   & 0.25   & 0.00 \\
			Cov($y_0$, $x$)  & 0.5   & 0.50 & 0.94 \\
			Cov($y_1$, $x$)  & 0.5   & 0.50   & 0.95\\\hline
			SPC $\rho_{partial} = 0.73$       &      &      &  \\\hline
			E($y_0$)       & 0.0     & 0.00     & 0.97 \\
			E($y_1$)       & 1.0     & 1.00     & 0.95 \\
			Var($y_0$)     & 1.0     & 1.00 & 0.94 \\
			Var($y_1$)     & 1.0     & 1.00 & 0.95 \\
			Cov($y_0$, $y_1$) & 0.8   & 0.80   & 1.00 \\
			Cov($y_0$, $x$)  & 0.5   & 0.50 & 0.95 \\
			Cov($y_1$, $x$)  & 0.5   & 0.50   & 0.94\\\hline
			SPC $\rho_{partial} = 0.99$       &      &      &  \\\hline
			E($y_0$)       & 0.0     & 0.00     & 0.95 \\
			E($y_1$)       & 1.0     & 1.00     & 0.95 \\
			Var($y_0$)     & 1.0     & 1.00 & 0.95 \\
			Var($y_1$)     & 1.0     & 1.00 & 0.94 \\
			Cov($y_0$, $y_1$) & 0.8   & 0.99   & 0.00 \\
			Cov($y_0$, $x$)  & 0.5   & 0.50 & 0.95 \\
			Cov($y_1$, $x$)  & 0.5   & 0.50   & 0.95
		\end{tabular}
		\caption{Parameter estimates for SPC with different partial correlations}
		\label{tab2}
	\end{table} 
	Table \ref{tab2} shows all statistics relevant to the potential outcomes. Statistics involving only one potential outcome are unbiased and have valid coverage rates, which means that even with incorrect specified partial correlation, we could derive plausible marginal distribution of potential outcomes. Since the partial correlation is set before imputation and there is no information about the correlation between potential outcomes in the data, we get valid inference for the marginal correlation between potential outcomes only when we specified the partial correlation correctly.  
	
	\section{Application}
	\label{sec:6}
	We apply the SPC algorithm to evaluate the effects of two different therapies on slowing the progression of HIV disease. The data comes from a study comparing the effects of four therapies (zidovudine alone, didanosine alone, zidovudine plus didanosine and zidovudine and zalcitabine) on preventing the deterioration of disease in adults with HIV-1 infected patients \cite{hammer1996trial}. For simplicity, we name them treatment A(zidovudine alone), B(didanosine alone), C(zidovudine plus didanosine) and D(zidovudine and zalcitabine). The data named ACTG175 is accessible in package \texttt{speff2trial} in \texttt{R}. We restrict our analyses to treatments A and B and perform out-of-sample prediction of hypothetical effect of treatments A and B for the remaining patients that were allocated to treatments C and D. Hammer et al. concluded that treatment with didanosine is superior to treatment with zidovudine. However, the overall treatment effect was found to be insufficient to recommend the therapy to a patient, a situation that is common in many medical interventions. 
	
	A total of 693 HIV-1 infected adults with CD4 cell counts in the range of 200 to 500 per cubic millimetre were randomised into the control (N = 316) and treatment (N = 377) group, while 670 patients were treated as out-of-sample. Fifteen baseline covariates are included which assess gender, age, weight, Karnofsky score, risk factors, prior antiretroviral therapy, CD4 cell count, and CD8 T cell count. We are interested in the number of days until the first occurrence of: 1) a decline in CD4 cell count of at least 50 2) an event indicating progression to AIDS, or 3) death. The larger number of days yields a more beneficial treatment effect. The individual treatment effect is defined as the number of days under treatment B minus the number of days under treatment A. We select the value of partial correlation as 0 and 0.7 to perform the sensitivity analysis so that the result yields distinct differences. 
	
	\begin{figure}[ht!]
		\begin{center}
			\resizebox{\textwidth}{!}{
				\subfigure[SPC $\rho_{partial} = 0$]{
					\label{boxplot:i}
					\includegraphics[scale=.5]{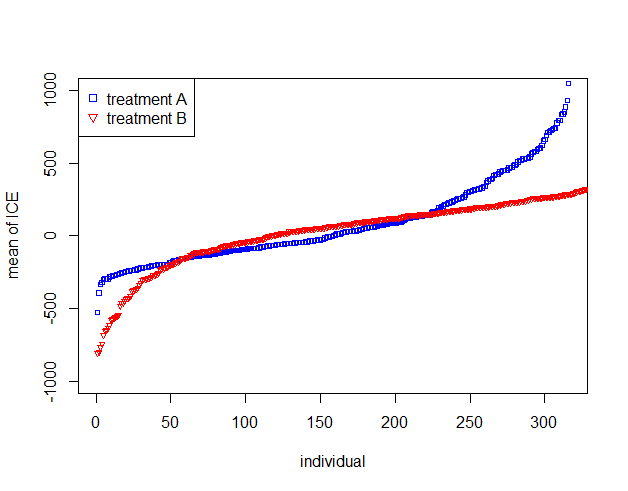}
				}
				\subfigure[SPC $\rho_{partial} = 0.7$]{
					\label{boxplot:j}
					\includegraphics[scale=.5]{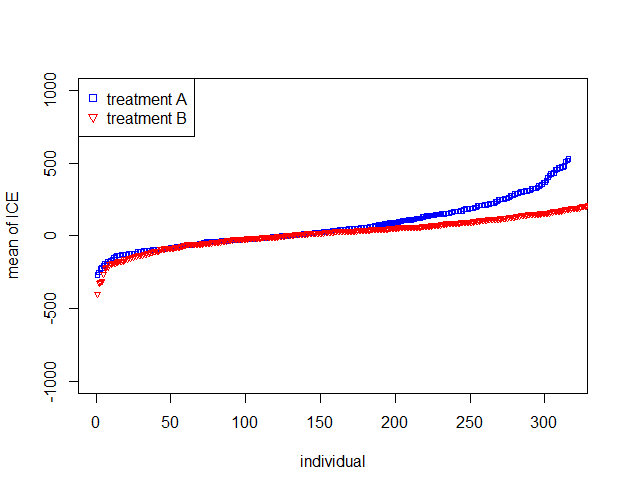}
				}
			}\\
		\end{center}
		\caption{Means of individual treatment effects for patients under treatment A and treatment B, which are arranged into ascending order.}
		\label{fig7}
	\end{figure}
	
	\begin{figure}[ht!]
		\begin{center}
			\resizebox{\textwidth}{!}{
				\subfigure[SPC $\rho_{partial} = 0$]{
					\label{boxplot:k}
					\includegraphics[scale=.5]{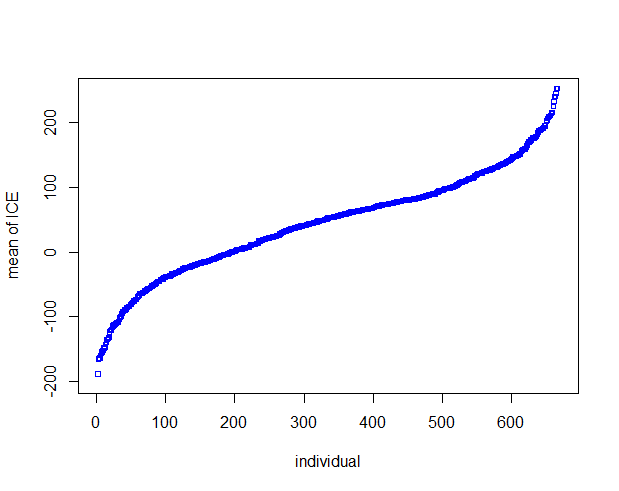}
				}
				\subfigure[SPC $\rho_{partial} = 0.7$]{
					\label{boxplot:l}
					\includegraphics[scale=.5]{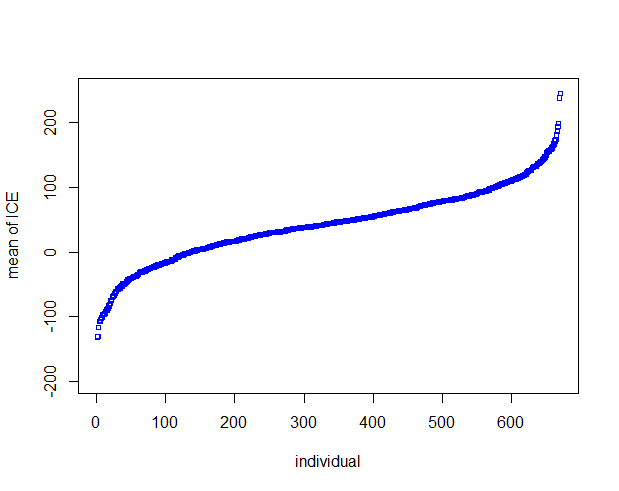}
				}
			}\\
		\end{center}
		\caption{Means of individual treatment effects for out-of-sample patients under treatment C and D, which are arranged into ascending order.}
		\label{fig8}
	\end{figure}
	Figure \ref{fig7} shows the results of individual treatment effects under treatment A and treatment B group with partial correlation specification 0 and 0.7. As expected, this approach could detect the heterogeneity of treatment effects. A large proportion of patients in the sample, whose individual treatment effects are larger than 0, are recommended to receive a treatment regimen with didanosine. However, treatment with zidovudine still yields greater clinical benefit for a fraction of units, whose individual treatment effects are smaller than 0. Since all covariates are balanced under two groups, the distributions of expected individual treatment effects under two treatments (A and B) are more similar when specifying a 0.7 partial correlation. Furthermore, the range of expected value of individual treatment effects is smaller, with a partial correlation of 0.7. The larger the partial correlation we set, the more convinced that all effect modifiers are included, and effects are identical across persons. Figure \ref{fig8} shows variability in individual effects in out-of-sample patients, which implies that our method could also be applied to a prediction scenario. 
	\begin{figure}[ht!]
		\begin{center}
			\resizebox{\textwidth}{!}{
				\subfigure[]{
					\label{boxplot:m}
					\includegraphics[scale=.5]{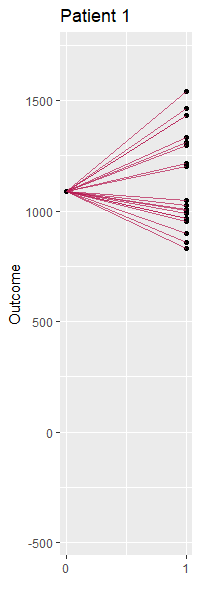}
				}
				\subfigure[]{
					\label{boxplot:n}
					\includegraphics[scale=.5]{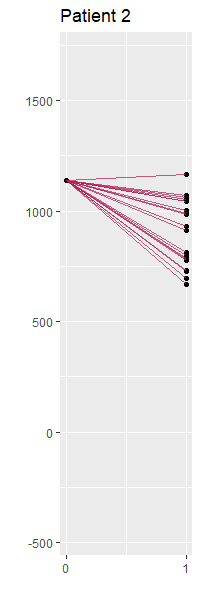}
				}
				\subfigure[]{
					\label{boxplot:o}
					\includegraphics[scale=.5]{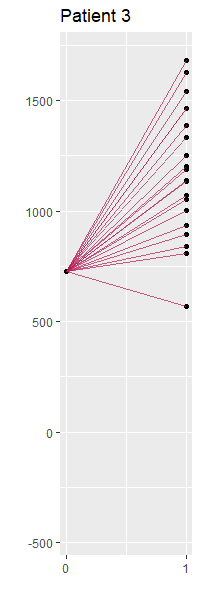}
				}
				\subfigure[]{
					\label{boxplot:p}
					\includegraphics[scale=.5]{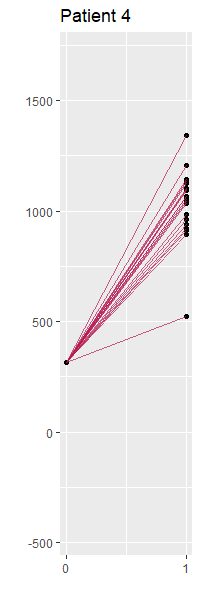}
				}
				\subfigure[]{
					\label{boxplot:q}
					\includegraphics[scale=.5]{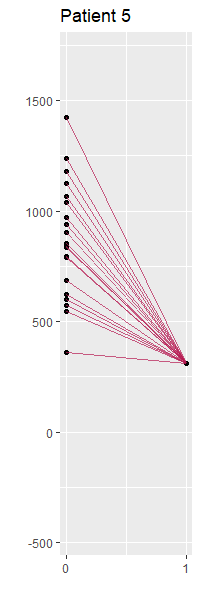}
				}
				\subfigure[]{
					\label{boxplot:r}
					\includegraphics[scale=.5]{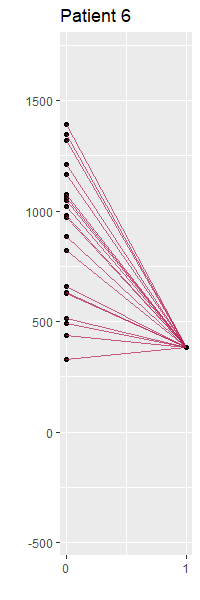}
				}
				\subfigure[]{
					\label{boxplot:s}
					\includegraphics[scale=.5]{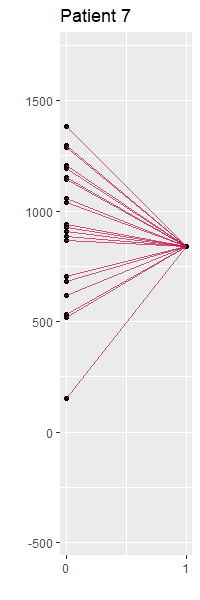}
				}
			}\\
		\end{center}
		\caption{Fan plot of Observed and imputed (m = 20) outcomes under treatment A (0) and B (0). The partial correlation is 0.}
		\label{fig9}
	\end{figure}
	
	\begin{figure}[ht!]
		\begin{center}
			\resizebox{\textwidth}{!}{
				\subfigure[]{
					\label{boxplot:t}
					\includegraphics[scale=.5]{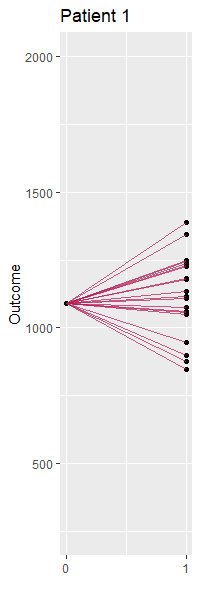}
				}
				\subfigure[]{
					\label{boxplot:u}
					\includegraphics[scale=.5]{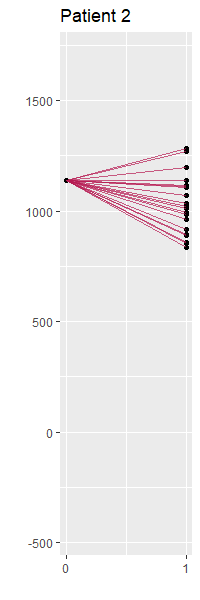}
				}
				\subfigure[]{
					\label{boxplot:v}
					\includegraphics[scale=.5]{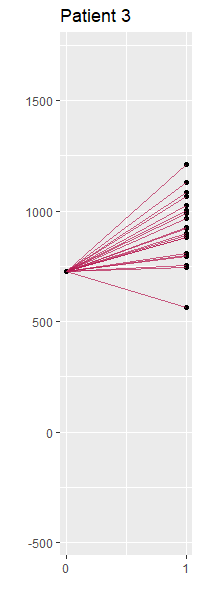}
				}
				\subfigure[]{
					\label{boxplot:w}
					\includegraphics[scale=.5]{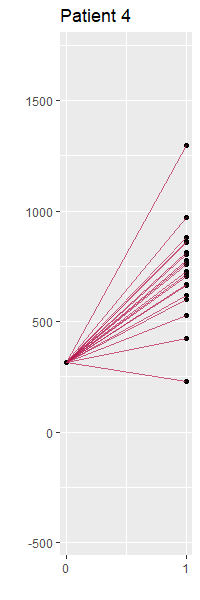}
				}
				\subfigure[]{
					\label{boxplot:x}
					\includegraphics[scale=.5]{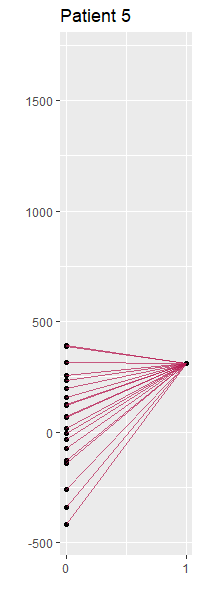}
				}
				\subfigure[]{
					\label{boxplot:y}
					\includegraphics[scale=.5]{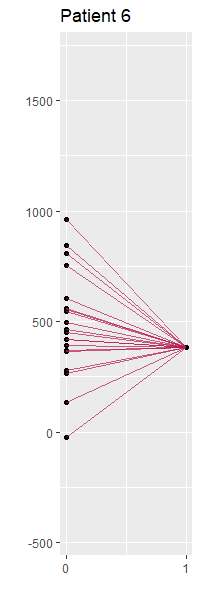}
				}
				\subfigure[]{
					\label{boxplot:z}
					\includegraphics[scale=.5]{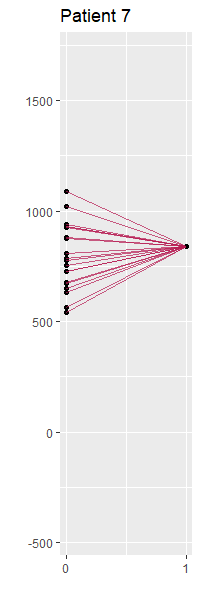}
				}
			}\\
		\end{center}
		\caption{Fan plot of Observed and imputed (m = 20) outcomes under treatment A (0) and B (0). The partial correlation is 0.7.}
		\label{fig10}
	\end{figure}
	Figure \ref{fig9} and \ref{fig10} display imputations by selected patients for two different values of partial correlation, 0 and 0.7. Each panel contains the observed outcome for the patient and $m = 20$ imputed values for the missing outcome. The imputed outcomes are sensitive to different values of partial correlation. Patient 5 benefits from treatment A when the partial correlation equals 0, while we derive the opposite conclusion when specifying the partial correlation as 0.7. The location of imputed outcomes is different under various partial correlations, and the scale shrinks when the partial correlation tends to 1.  
	
	\section{Discussion}
	\label{sec:7}
	We propose a multiple imputation approach to replace missing outcomes with plausible values to estimate the individual treatment effect. Treatment assignment is currently steered by the average treatment effect. This may lead to suboptimal individual treatment decisions because the treatment effects are assumed to be homogeneous. We have demonstrated that the proposed SPC algorithm allows for the imputation of heterogeneous treatment effects under a given partial correlation. 
	
	The sensitivity analysis of the partial correlation between potential outcomes in section 5.2 demonstrates that different values of the partial correlation yield similar results in terms of marginal distributions of potential outcomes and the average treatment effect. However, the closer the specified partial correlation is to the \emph{true} value, the less biased the estimated individual treatment effect will be. Since one cannot obtain information about the partial correlation from the observed data, the determination of the partial correlation should be set to a plausible range based on previous investigations or expert knowledge. In addition, it may be useful to perform a sensitivity analysis to see how imputed outcomes differ with different partial correlations. 
	
	An advantage of our multiple imputation approach to individual treatment effects is that it provides an estimate of the uncertainty of the imputed outcomes and hence, of the individual treatment effects. One could obtain the posterior distribution of the individual treatment effects for a unit from multiply imputed datasets, from which we can learn the probability of benefit from the treatment at the individual level. Since we incorporate the partial correlation when imputing, researchers could apply complete-data analyses to explore potential variables. This accounts for residual heterogeneity of treatment effects or additional effect modifiers. 
	
	In our illustration of the SPC algorithm, we applied Bayesian imputation under the normal linear model. It is possible to use other imputation techniques (parametric or non-parametric imputation methods). The behaviour of such methods has not yet been studied. Since SPC is a hybrid of FCS and JM, it will provide valid inferences on data that are missing at random.  Another useful property of SPC is that under the assumption of ignorable treatment assignment, researchers can skip explicit modelling of the probability of assignment. 
	
	In the application study, we focus on the comparison of treatments A and B. It is possible to generalise the multiple treatment comparison (treatment A, B, C, and D). By imputing unobserved outcomes, we could then recommend the optimal treatment to each unit among four treatments. One could benefit from our method when performing an experiment that has been investigated on a different population. Some proven effect modifiers may be difficult to collect when performing the same experiment in other regions or countries. In such a case, the inference would include the heterogeneity of treatment effects explained by the uncollected factors by specifying a reasonable partial correlation.         
	
	This is an initial study on MI to the individual treatment effect. The simulation study used a basic randomised trial with a correctly specified imputation model. Further work should be done to extend discrete and semi-continuous outcomes. Another challenge is to develop imputation techniques for studies that collect post-treatment variables. All in all, we believe that our methodology for incorporating the partial correlation represents an important advance in estimating individual treatment effects. We hope that our method may attribute to the growing body of work on personalised statistics and individual treatment effects. 
	\newpage
	\bibliographystyle{apacite}
	\bibliography{MIICE}    
\end{document}